\documentclass[12pt]{article}

\onecolumn
\usepackage{graphicx}

\newtheorem{theorem}{Theorem}
\newtheorem{lemma}{Lemma}

\newtheorem{definition}{Definition}

\usepackage{graphicx}

\begin{document}

\date{}

\title{Routing for Global Congestion Avoidance}

\author{
\Large  \hspace*{7mm} Zohre R. Mojaveri and Andr\'as Farag\'o\\
Department of Computer Science \\
The University of Texas at Dallas \\
Richardson, Texas\\
}

\maketitle
       \thispagestyle{empty}

\section{Introduction}

Traditionally, it is desired that a route in a network uses lightly loaded links, to reduce blocking probability. 
A usual way to implement this goal is to assign a weight to each link, such that it reflects the traffic load on the link. Having 
these weights, we can look for a minimum weight path. The latter can be found by classic methods, such as Dijkstra's algorithm. 

The link weight can be derived from various practical parameters that are related to the load. 
Examples are delay, queue length in packet switching networks,
blocking probability or the number of occupied circuits in circuit switched networks, etc.

While the above approach aims at {\em locally} avoiding heavily loaded links, it only takes into account the congestion of {\em individual links}
(this is what we refer to as local view). It is not sensitive to congestion in a whole subnetwork, which would be a more globally oriented view.
For example, it does not distinguish a link, which is overloaded only in itself, from another one, which is overloaded together with its whole 
neighborhood, being part of a congested subnetwork. 

Our goal is to study routing strategies for {\bf\em global congestion avoidance}, that is, finding routes that avoid 
congested {\bf\em subnetworks,} not only congested {\bf\em links.} Specifically, we address the issues outlined in the next sections.

\section{Modeling Congested Subnetworks}

{\bf Graphs.} We model the network as a graph, in which the edges represent the network links.  
We consider undirected simple graphs (no self-loops and multiple edges). Such a graph is given 
as $G=(V,E)$, where $V$ is the set of nodes (vertices), and $E$ is the set of edges. Sometimes the notations $V(G)$ and $E(G)$ are used,
if we need to specify that the nodes/edges belong to which graph. When we refer to a subgraph (which may be a path), 
then often we simply identify it with its set of nodes. For example, if $P$ is a path and $S$ is a subgraph, then $P\cap S=\emptyset$ 
means that the path $P$ avoids the subgraph $S$.

\subsection{Congested Links}
\label{links}

\begin{itemize}

\item Each link (edge) is characterized by its traffic load, relative to the link capacity. For example, a link may be 60\% loaded, that is,
60\% of the link capacity is being used; then we assign a weight of 0.6 to this link. 

\item We set a link congestion threshold. Whenever the weight of the link is above the threshold, we call it congested. 
For example, if the threshold is 0.7, then a link with 60\% relative load is not congested, but another one with relative load of 80\% is. 

\item Network filtering: keep only those edges of the graph that represent congested links, according to the link congestion threshold. 
Let us call the arising graph the {\bf\em congested core} of the network.

\end{itemize}

\subsection{Congested Subnetworks}

In one sentence, a {\bf\em congested subnetwork is defined as a dense subgraph of the congested core of the network.}
The congested core has been already defined in Section~\ref{links}; we explain in the next subsection what it means that  a subgraph is considered dense. But before going into that, let us mention that our approach combines two facets of congestion:
\vspace*{-2mm}
\begin{itemize}

\item {\bf\em Traffic congestion:} it is what we explained in Section~\ref{links}.

\item {\bf\em Topological congestion:} this is what is captured by dense subgraphs. The density refers to the {\em topology} of the subnetwork, not to the 
traffic. On the other hand, these dense subgraphs are only sought within the congested core of the network, which is defined on the basis of traffic
(see Section~\ref{links}). This way we integrate the traffic congestion and topological congestion aspects.

\end{itemize}

\subsubsection{Graph Density Measures}

We start with the most general setting, then narrow it down to  meaningful specific cases.

\begin{definition} {\bf (Edge-monotone function)}
Let $f(G)$ be a function that assigns a number to any graph. We say that the function is edge-monotone, if adding more edges to the graph cannot 
decrease the value of the function. 
\end{definition} 

\begin{definition} {\bf (Graph density measure)}
A function $\rho(G)$ is called a {\bf\em graph density measure,} if the function is edge-monotone, and  $\rho(G)\geq 0$ holds for any graph $G$. 
The value $\rho(G)$ is called the {\bf\em density} of the graph $G$. 
\end{definition}

\noindent
{\em Note:} If we are interested in the density of a subgraph $S$, then we use the notation $\rho(S)$.

As we are going to deal with different $\rho$ functions,  the various types of densities are distinguished by referring to $\rho$-density. For example,
if $\rho_1,\rho_2$ are two such functions, then they give rise to the $\rho_1$-density and the $\rho_2$-density of the graph. These densities may be different. 
Occasionally they have special names, such as edge-density, etc.

\subsubsection{Examples of Graph Density Measures}
\label{examples}

Some interesting density measures are listed below. 

\begin{itemize}

\item 
\noindent {\bf Edge density.}
$$ \rho_1(S)=\frac{|E(S)|}{|V(S)|}\,.$$
Observe that $|E(S)|$ is precisely twice the sum of the degrees in the subgraph $S$, since each edge contributes 2 to 
the sum of degrees. Therefore, we can write
$$\rho_1(S)=\frac{1}{2}\frac{\sum_{v\in S} d_S(v)}{|V(S)|}=\frac{1}{2}\: \overline d(S),$$
where $d_S(v)$ is the degree of vertex $v$ within the subgraph $S$, and $\overline d(S)$ is the {\em average degree} in $S$.
Thus, $\rho_1(S)$ is precisely half of the average node degree in $S$.

\item 
\noindent {\bf $k$-core.} Let us use the density measure 
$$\rho_2(S)=\min \{d_S(v)\;|\; v\in S\}.$$
Then a subgraph with $\rho_2(S)\geq k$ is a subgraph in which each node has degree at least $k$, within the subgraph. For a given $k$,
the largest such subgraph (where the size is measured by the number of nodes)  is called the $k${\bf\em -core} of the graph. The $k$-core in any graph is unique. 
The reason is that if two subgraphs both have the property that each degree is at least $k$, then their union also has this property, leading to a unique
maximum subgraph with the property.

\item {\bf $k$-clique density.} A $k$-clique is a complete graph with $k$ vertices. 
For a fixed $k$, define the density measure as 
$$\rho_3(S)\;=\; \frac{\mbox{number of}\; k\mbox{-cliques in}\; S}{|V(S)|}$$
If $k=2$, this gives back the edge density $\rho_1(S)$. For $k>1$ it is generally different. 

\item {\bf Squared degree density.} Let us use the density measure 
$$\rho_4(S) \;=\; \frac{\sum_{v\in S} d_S^2(v)}{|V(S)|}\,.$$
This is similar in spirit to $\rho_1(S)$, which is proportional to the sum of degrees, except that 
$\rho_4(S)$ is the sum of squared degrees. Consequently, $\rho_4(S)$ is more reflective of the large degrees. 

\item {\bf Edge-connectivity.}
Let $\lambda(S)$ denote the edge connectivity of the subgraph $S$, that is the minimum number of edges whose deletion disconnects the graph.
Let $$\rho_5(S)\;=\; \lambda(S).$$
Maximizing this density measure means finding a subgraph with the highest edge-connectivity. 

\end{itemize} 

It holds for all the above density measures that a subgraph of maximum $\rho$-density can be found by (different) polynomial time algorithms. That is,
the {\em $\rho$-densest subgraph problem} can be solved efficiently for all the above densities. 
There are, however, other measures for which this maximization is {\bf NP}-hard. See many more examples and related results in \cite{algo}.

\section{Routes that Avoid All Dense Subgraphs: Global Congestion Avoidance}

Let us now define the property of the path (route) we are looking for.

\subsection{Base Version}

\begin{definition} \label{CAP} {\bf (Congestion-Avoiding Path (CAP))}
Let $G$ be a graph, representing a network, and let $C(G)$ be its congested core (see {\em Section~\ref{links}}). Let $s,t\in V(G)$ be two distinct 
nodes. We say that an $s-t$ path $P$ is a {\em Congestion-Avoiding Path (CAP)} between $s$ and $t$, 
with respect to a graph density measure $\rho$, if $P\cap S =\emptyset$ holds for every subgraph $S\subseteq C(G)$, which is densest within $C(G)$ 
with respect to $\rho$.
\end{definition}

\noindent
{\bf Interpretation:} An $s-t$ CAP is an $s-t$ path that avoids all densest subgraphs within the congested core. The density is meant with respect to 
the density measure $\rho$.

\subsection{A Stronger Version}

We may require more: avoiding not only the {\em densest} subgraphs of the congested core, but all subgraphs that have $\rho$-density at least 
a given value $\rho_0$.

\begin{definition} {\bf (Density index of a path)}
Let $P$ be a path in a graph $G$, and let $\rho$ be a graph density measure.
The {\em density index} of $P$, denoted by $\rho'(P)$,  is defined as the smallest number $\rho_0$, for which it holds that $P$ avoids every subgraph 
$S$ of $C(G)$ with $\rho(S)\geq \rho_0$, where $C(G)$ is the congested core. In formula,
$$\rho'(P)\;=\; \min\{\rho_0\;|\; \forall S\subseteq C(G):\; \rho(S)\geq \rho_0 \Rightarrow P\cap S=\emptyset\}
$$
\end{definition}

\noindent
Now we can define our target algorithmic task:

\begin{center}
\begin{tabular}{|l|}
\hline
{\sc Path with prescribed density index} \\[3mm]

{\bf Input:} Graph $G$ with congested core $C(G)$, two distinct  nodes  $s,t\in V(G)$, \\ 
     \hspace*{13mm}        density measure $\rho$,  and a real number  $\rho_0\geq 0$.\\[2mm]

{\bf Task:}\,\, Find an $s$-$t$  path $P$ in $G$, such that $\rho'(P)\geq \rho_0$, \\ 
\hspace*{13mm} or else declare that no such path exists. \\
\hline
\end{tabular}
\end{center}

\noindent {\bf Interpretation:} We are looking for a path that connects $s$ and $t$ (the source and target nodes), 
such that the path avoids {\em every} subgraph $S$ with $\rho(S)\geq \rho_0$. That is, every subgraph is avoided that has
$\rho$-density at least $\rho_0$. By choosing $\rho_0=\max_{S\subseteq C(G)} \rho(S)$, we get back the base case 
of $s-t$ CAP introduced in Definition~\ref{CAP}.

If we carry out this task in the congested core of the network, then we can indeed implement global congestion avoidance,
since the path will avoid {\em all} congested subnetworks. Recall that the latter are defined as dense subgraphs of the congested core 
of the network (according to the chosen density measure).

\section{Algorithmic Issues}
Having defined global congestion avoidance, the next natural question is: how to solve the above outlined task {\em algorithmically}, that is,
how to find a path with prescribed density index?

Observe that even if we use an ``easy" density measure (for which the densest subgraph can be found in polynomial time, see the examples in
Section~\ref{examples}), it is still not clear 
how to find an $s$-$t$  path that avoids {\em all} subgraphs with $\rho(S)\geq \rho_0$. Note that there may be exponentially many such subgraphs, 
so listing all of them is generally not feasible.

Nevertheless, we can prove that at least in some cases the problem is solvable in polynomial time. First we need a definition:

\begin{definition} \label{eff} {\bf (Efficient graph density measure.)} 
We say that a graph density measure $\rho$ is {\bf efficient,} if there exists a polynomial time algorithm, which, for any graph $G$ and 
parameter $\rho_0\geq 0$,  can list all subgraphs $S$ in  $G$, such that $\rho(S)\geq \rho_0$ holds.
\end{definition}

Now we can state our general theorem. It is easy to prove, but we still state it as a theorem, because it makes possible to solve the task 
{\sc Path with prescribed density index} in polynomial time, whenever an efficient graph density measure is used.

\begin{theorem} \label{thm1} For any efficient graph density measure $\rho$ and parameter $\rho_0\geq 0$, the problem 
{\sc Path with prescribed density index}  can be solved in polynomial time.
\end{theorem}

\noindent {\bf Proof.} Since the graph density measure is assumed efficient, therefore, according to Definition~\ref{eff}, there exits a 
polynomial time algorithm, which, for any graph $G$ and parameter $\rho_0\geq 0$,  can list all subgraphs $S$ in  $G$, such that $\rho(S)\geq \rho_0$ holds.
Let us call this algorithm $\cal A$, and run it on the congested core  $C(G)$. Let $S_1,\ldots,S_k$ be the obtained list of subgraphs. These are all the subgraphs
in $C(G)$, for which $\rho(S_i)\geq \rho_0$ holds, $i=1,\ldots,k$. Observe that the value of $k$ must be polynomially bounded, since $\cal A$ runs in 
polynomial time, therefore, it can output at most a polynomial number of subgraphs.

Now let $G'$ be the graph obtained by removing all the subgraphs $S_1,\ldots,S_k$ from $C(G)$. This can clearly be carried out in polynomial time, since 
$k$ is bounded by a polynomial. Then we can define the output, as follows:
\begin{enumerate}

\item If at least one of $s$ and $t$ is not in $G'$, then output the message ``{\sc there is no $s-t$ path with $\rho'(P)\geq \rho_0$.}"

\item If both $s$ and $t$ are in $G'$, then find a shortest  $s-t$ path in $G'$, if any. If such a path is found, the output it as the result.
Otherwise, output the message ``{\sc there is no $s-t$ path with $\rho'(P)\geq \rho_0$.}"

\end{enumerate}
The shortest path in 2 can be found by Dijkstra's Algorithm in polynomial time; it also detects if no such path exists. 
The correctness of the result follows from the fact that an $s-t$ path with $\rho'(P)\geq \rho_0$ can only exist if there is an 
$s-t$ path that avoids {\em all} subgraphs with $\rho(S)\geq \rho_0$. Since these are precisely the subgraphs $S_1,\ldots,S_k$,
listed by $\cal A$, therefore, our algorithm indeed outputs the correct result, whenever an $s-t$ path with $\rho'(P)\geq \rho_0$ exists.

\hfill $\clubsuit$

\section{Efficient Graph Density Measures}

In view of Theorem~\ref{thm1}, it is natural to ask: which graphs density measures are efficient in the sense of Definition~\ref{eff}?
Note that it requires that we can list all subgraphs $S$ with $\rho(S)\geq \rho_0$, for any $\rho_0$, in polynomial time. Therefore, it is not
sufficient if we can just find a {\em densest} subgraph in polynomial time. (The latter condition is satisfied by each density measure in
Section~\ref{examples}.)

There are (at least) two among  the density measures in
Section~\ref{examples} for which we can prove they are efficient.

\begin{itemize}

\item $k${\bf -core.} The efficiency of $k$-core follows from two facts. (1) It can be found in polynomial time for any $k$, see \cite{algo}.
(2) The $k$-core for any $k$  is unique. The reason for the latter is that if two subgraphs both have the property that each degree is at least $k$, then their union also has this property, leading to a unique maximum subgraph with the property.

\item {\bf Edge connectivity}. This density measure characterizes the graph density by the edge connectivity of a subgraph that has maximum edge connectivity. 
It can be found in polynomial time, see \cite{algo}. However, to precisely satisfy Definition~\ref{eff},  we need a polynomial time  algorithm that can list 
{\em all} subgraphs that have edge connectivity at least $\rho_0$.  Note that here $\rho_0$ may be smaller than the largest edge connectivity 
that occurs in a subgraph. Unfortunately, in general, this cannot be done in polynomial time. For example, if we consider a complete graph on $2n$ nodes,
and take $\rho_0=n$, then every complete subgraph on $\rho_0+1=n+1$ nodes will be $\rho_0$-connected. Since there are exponentially many such subgraphs
in this case, they cannot be all listed in polynomial time.  

On the other hand, we can still satisfy efficiency in the somewhat weaker sense of Definition~\ref{CAP}. The reason is that if we look for maximal size 
subgraphs with maximum edge-connectivity (they can be found in polynomial time, see \cite{algo}), there can only be linearly many of them. The reason is 
that any two such maximal subgraphs, if they do not coincide, then they are {\em node-disjoint.} This follows from the fact that the union of any two 
overlapping $k$-edge-connected subgraphs remain edge connected, as we prove below in Lemma~\ref{union}. Therefore, due to the node-disjointedness,
the summed size of the maximal $k$-edge-connected subgraphs cannot be more than the number of nodes in the whole graph.

\end{itemize}

\begin{lemma}\label{union} {\bf (Union of overlapping $k$-edge-connected subgraphs)} Let $G$ be a graph, and $G_1, G_2$ be two 
$k$-edge-connected subgraphs in $G$. If $V(G_1)\cap V(G_2)\neq \emptyset$, then $G_1\cup G_2$ is also a $k$-edge-connected subgraph of $G$.
\end{lemma}

\noindent
{\bf Proof.} Let $w\in V(G_1)\cap V(G_2)$. Pick a node $u\neq w$ in $G_1$, and a different node $v\neq w$ in $G_2$.
We show that there exist $k$ edge-disjoint paths connecting $u$ and $v$ in $G_1\cup G_2$. Assume indirectly it is not true. Then, by Menger's 
Theorem\footnote{See any advanced textbook on graph theory, such as \cite{diestel}.} there are is a cut of at most $k-1$ edges separating $u$ and $v$  
in $G_1\cup G_2$. We show it leads to a contradiction. Let $C$ be such a cut. Then $C$ cannot separate $u$ and $w$ in $G_1$, since $G_1$ is $k$-connected. Therefore, after removing $C$, there is still a path $P_1$ in $G_1$ connecting $u$ and $w$. Similarly, $C$ cannot separate $w$ and $v$ in $G_2$, since $G_2$ is also $k$-connected, so after removing $C$, there is still a path $P_2$ in $G_2$ connecting $w$ and $v$. Then, after removing $C$,  
we can still reach $v$ from $u$ in $G_1\cup G_2$ by proceeding on
$P_1$ from $u$ to $w$, and then continuing on $P_2$ from $w$ to $v$. Thus, $C$ did not separate $u$ and $v$ in $G_1\cup G_2$, a contradiction. 
Thus, we conclude that $G_1\cup G_2$ must be a $k$-edge-connected subgraph of $G$.

\hfill $\clubsuit$

It is worth mentioning that not all graph density measures lead to a polynomially bounded number of densest subgraphs. 

\section{More About the $k$-core}

As we have seen above,  the density concept best suited for Global Congestion Avoidance is the $k$-core. Below we list some properties of 
the $k$-core that make it indeed an attractive density measure.

\begin{itemize}

\item As already mentioned, the $k$-core is {\em unique} for any $k$.

\item It is algorithmically easy to find it (see \cite{algo}).

\item The $k$ cores of a graph, for different values of $k$, form a nested hierarchy, as illustrated in Fig. ~\ref{fig1}

\begin{figure}[ht]
\centering
\hspace*{10mm}
\includegraphics[width=0.80\columnwidth]{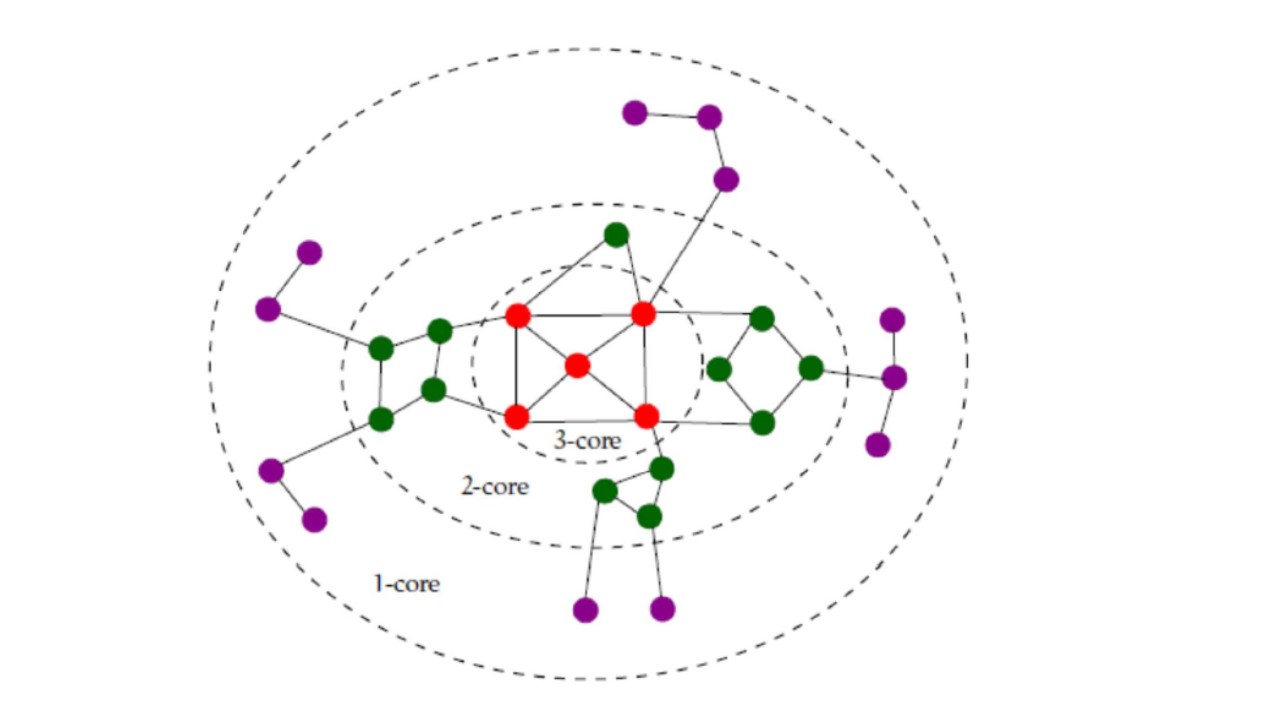}
\caption{$k$-core illustration}
\label{fig1}
\end{figure}

\item It is related to the often investigated degree distribution, see \cite{kong}.

\item A lot is known about the behavior of the $k$-core in random graphs, for a survey see, e.g.,  \cite{kong}.

\item The largest $k$ for which the graph has a non-empty $k$-core is a good approximation of the maximum clique size 
in many real-life graphs, see \cite{walteros}. Note that the maximum clique size is an {\bf NP}-complete problem, it is 
quite useful to have an easily computable approximation that is close to it in many real-life graphs.

\end{itemize}

\section{Creating New Efficient Graph Density Measures from Old Ones}

The next lemma shows a way to create new efficient graph density measures from existing ones.

\begin{lemma}\label{intersect}
If $\rho_1, \rho_2$ are efficient graph density measures (in the sense of Definition~\ref{eff}), then the measures
\begin{equation}\label{min}
\rho_{min}(S)\;=\; \min\{\rho_1(S),\rho_2(S)\}
\end{equation}
and 
\begin{equation}\label{max}
\rho_{max}(S)\;=\; \max\{\rho_1(S),\rho_2(S)\}
\end{equation}
also remain efficient.
\end{lemma}

\noindent {\bf Proof.} For any density measure $\rho$, let us call the list of graphs with $\rho(S) \geq \rho_0$ the $\rho_0$-dense list of $\rho$.
It follows from (\ref{min}) that the subgraphs for which $\rho_{min}(S) \geq \rho_0$ holds must have the property that both 
$\rho_1(S) \geq \rho_0$ and $\rho_2(S) \geq \rho_0$ are satisfied. Therefore, the subgraphs in the $\rho_0$-dense list of $\rho_{min}$ are precisely those
that are both on the $\rho_0$-dense list of $\rho_1$ and on the $\rho_0$-dense list of $\rho_2$. It means that $\rho_{min}$ is also efficient, and its
$\rho_0$-dense list is the intersection of the $\rho_0$-dense lists of $\rho_1$ and $\rho_2$.

Similarly, it follows from (\ref{max}) that the subgraphs for which $\rho_{max}(S) \geq \rho_0$ holds must have the property that at least one of  
$\rho_1(S) \geq \rho_0$ and $\rho_2(S) \geq \rho_0$ are satisfied. As a result, we obtain that $\rho_{max}$ is also efficient, and its
$\rho_0$-dense list is the union of the $\rho_0$-dense lists of $\rho_1$ and $\rho_2$.

\hfill $\clubsuit$

The property stated in Lemma~\ref{intersect} allows that we can efficiently  find routes that either simultaneously satisfy several global congestion avoidance properties, or satisfy at least one of several such properties. Note that (\ref{min}) and (\ref{max}) can be directly extended to the minimum or maximum
of more than two measures, and the Lemma still carries over to these cases. Applying it recursively also allows us to claim efficiency in more 
complicated cases. For example, if $\rho_1,\rho_2\rho_3,\rho_4$ are efficient graph density measures, then the measure
$$
\rho(S)\;=\; \min\{\; \max\{\rho_1(S),\rho_2(S)\},\; \max\{\rho_3(S),\rho_4(S)\;\}
$$
remains efficient. Even more generally, we can build any constant size expression from min and max operations, 
and the resulting density measure still remains efficient.

\section{Open Problems}
\label{open}

In conclusion, we list some open problems and conjectures in connection with our approach.

\begin{itemize}

\item For a number of density measures it is known that maximizing them is {\bf NP}-hard, but efficient approximation algorithms with proven approximation
guarantees are often available (see \cite{algo}). How can our framework be extended to such cases? 

\item Extend the framework for the case when we look for several {\bf\em disjoint} $s$-$t$ paths (i.e., they do not share any node other than $s,t$), such that 
all the paths have the global congestion avoidance property. 

\item Consider the above multi-path problem, but with the difference that we do not require all paths to satisfy the  
global congestion avoidance property, it is enough if one of them satisfies it (this one can serve as primary path, the others as backup paths).

\end{itemize}

\end{document}